\documentstyle[twoside,fleqn,espcrc2,psfig]{article}

\title{Complex structure of a DT surface with $T^{2}$ topology
\thanks{presented by N.Tsuda}}

\author
{
H.Kawai$^{\,\, {\rm a}}$, N.Tsuda \address{
National Laboratory for High Energy Physics (KEK), 
Tsukuba 305, Japan}
and T.Yukawa$^{\,\, {\rm a,}}$ \address{
Coordination Center for Research and Education, 
The Graduate University for Advanced Studies, Hayama-cho, 
Miura-gun, Kanagawa 240-01, Japan}
}

\begin{document}
\begin{abstract}
A method of defining the complex structure($i.e.$ moduli) for dynamically 
triangulated(DT) surfaces with torus topology is proposed.
Distribution of the moduli parameter is measured numerically and compared 
with the Liouville theory for the surface coupled to $c = 0, 1$ and $2$ 
matter.
Equivalence between the dynamical triangulation and the Liouville theory 
is established in terms of the complex structure.
\end{abstract}
\maketitle

\section{Introduction}
Two-dimensional quantum gravity(2DQG) is interesting from the point of view 
of string theory in non-critical or critical dimensions and the statistical 
mechanics of random surfaces.
As has been known the matrix model and the continuous theory, {\it i.e.} the 
Liouville field theory, exhibit the same critical exponent and Green's 
function.
However, the mutual relation between DT and the Liouville theory has not 
been so evident, because the DT surfaces are known to be fractal and its 
typical fractal dimension is considered to be four.
On the other hand, we will show in this paper, the complex structure is 
well defined even for the DT surfaces, and we can check a precise equivalence 
of the DT and the Liouville theory.

It is known that 2DQG can be interpreted as a special case of the 
critical string with the conformal mode having a linear dilaton background.
The critical string is defined to have two local symmetries, 
$i.e.$ the local scale(Weyl) invariance and the reparametrization($Diffeo$) 
invariance.
After imposing these two symmetries on the world-sheet, no degree of freedom 
is left for the metric $g_{\mu \nu}$ except for a set of parameters $\tau$, 
which specify the moduli space ${\cal M}$ of the complex structure: 
$
{\cal M} = \{ g_{\mu \nu} \} / Diffeo \otimes Weyl = \{ \tau \}.
$
Therefore, if we find a way to impose the local scale invariance on DT, we 
achieve a step closer to the study of the numerical simulations of the critical 
string, although it is not very easy at the present stage.
However, considering the complex structure is very useful for obtaining 
clear signals in various measurements because the complex structure can be 
extracted independently to the rather complicated fluctuations of the 
conformal mode.
In our previous work \cite{Cmplx_Struc}, we have established how to define 
and measure the complex structure and the conformal mode on the DT surfaces 
with $S^{2}$ topology.
To be concrete we focus on the case of torus $T^{2}$ in this study, but the 
generalization would be straightforward.
%
\section{Determination of a moduli on the torus}
\begin{figure}
\vspace*{0cm}
\centerline{\psfig{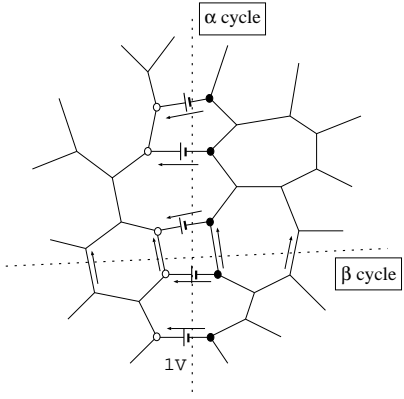}} 
\vspace{-1.0cm}
\caption
{
Configuration of batteries inserted across the $\alpha$-cycle.
The vertices with open circle denote the $v_{left}$ and with close 
circle denote the  $v_{right}$, respectively. 
}
\label{fig:Net_current}
\vspace{-0.6cm}
\end{figure}

In the continuous formulation the period $\tau$ is obtained by the following 
procedure. 
First we introduce harmonic $1$-form $j_{\mu}dx^{\mu}$ where $j_{\mu}$ 
satisfies the divergence and rotation free conditions 
$\partial_{\mu}j^{\mu} = 0$ and $\partial_{\mu}j_{\nu} - 
\partial_{\nu}j_{\mu} = 0$ respectively, with 
$j^{\mu} = \sqrt{g} g^{\mu \nu} j_{\nu}$.
Since there are two linearly independent solutions, we can impose 
two conditions such as
\begin{equation}
\oint_{\alpha} j_{\mu} dx^{\mu} = 0, \; 
\oint_{\beta} j_{\mu} dx^{\mu} = \frac{1}{r},
\label{eq:pot_cond}
\end{equation}
where $\alpha$ and $\beta$ represent two independent paths on the torus 
which intersect each other only once and $r$ denotes the resistivity of 
the surface. 
Under these conditions the period $\tau$ is given by 
\begin{equation}
\tau \equiv \frac{\oint_{\alpha} j_{\mu} dx^{\mu} + i\oint_{\alpha} 
\tilde{j}_{\mu} dx^{\mu}}{\oint_{\beta} j_{\mu} dx^{\mu} + i\oint_{\beta} 
\tilde{j}_{\mu} dx^{\mu}} 
= \frac{i r \oint_{\alpha} \tilde{j}_{\mu}dx^{\mu}}
{1 + i r \oint_{\beta} \tilde{j}_{\mu} dx^{\mu}},
\end{equation}
where $\tilde{j}_{\mu}$ is the dual of $j_{\mu}$ defined by 
$\tilde{j}_{\mu} = \epsilon_{\mu \nu} \sqrt{g} g^{\nu \lambda} j_{\lambda}$.
This procedure can be easily translated to the case of triangulated 
surfaces by identifying $j_{\mu}$ with the current on the resistor network 
of the dual graph. 
Then $r \oint_{\alpha} j_{\mu} dx^{\mu}$ and 
$\oint_{\alpha} \tilde{j}_{\mu} dx^{\mu}$ correspond to the potential drop 
along $\alpha$-cycle and the total current flowing across $\alpha$-cycle 
respectively.
We can easily impose two conditions eq.(\ref{eq:pot_cond}) by inserting 
electric batteries in the dual links crossing the $\alpha$-cycle and apply 
constant voltages(1V), see Fig.\ref{fig:Net_current}.
%
%
Writing the electric potential at the vertex $v$ as $V(v)$ and assuming 
that each bond has resistance $1 \Omega$, the current conservation reads 
$
V(v) = \frac{1}{3} \{ \sum_{1,2,3} V(v_{i}) + \delta_{\alpha\mbox{-cycle}} \},
$
where $\delta_{\alpha\mbox{-cycle}}$ represents the voltages of the 
batteries placed along the $\alpha$-cycle and $v_{1,2,3}$ are the three 
neighboring vertices of $v$.
We solve these set of equations iteratively by the successive 
over-relaxation method, and estimate the total currents 
flowing across the $\alpha$- and $\beta$-cycles as 
$
\oint_{\alpha} \tilde{j}_{\mu} dx^{\mu} = \sum_{{\tiny \alpha-\mbox{cycle}}} 
\{V(v_{right}) - V(v_{left}) + 1\}, \;\; \mbox{and} \;\; 
\oint_{\beta} \tilde{j}_{\mu} dx^{\mu} = \sum_{{\tiny \beta-\mbox{cycle}}} 
\{V(v_{right}) - V(v_{left})\}
$ 
respectively, where $V(v_{left})$ and $V(v_{right})$ denote the potentials 
of vertices placed at  either side of the $\alpha$- or $\beta$-cycle.
In the case of $T^{2}$ topology the resistivity is not easily determined 
because of the lack of the $SL(2,C)$-invariance as we have made use of in 
the $S^{2}$ topology case\cite{Cmplx_Struc}.
Here, we borrow the resistivities obtained in case of the $S^{2}$ topology 
in order to determine the moduli of the torus.

Here, we present the theoretical predictions of the distribution function of 
the moduli.
The genus-one partition function with c scalar fields is given by 
\cite{Torus_Part} by 
\begin{equation}
Z \simeq \int_{\cal F} \frac{d^{2}\tau}{(\tau_{2})^{2}} \; \{C(\tau)\}^{c-1},
\label{eq:Z_torus}
\end{equation}
\begin{equation}
C(\tau) = (\frac{1}{2} \; \tau_{2})^{-\frac{1}{2}} e^{\frac{\pi}{6} \; 
\tau_{2}} |\prod_{n=1}^{\infty} (1- e^{2\pi i \tau n}) |^{-2}, 
\end{equation}
where $\tau$ is a moduli parameter $\tau = \tau_{1} + i \tau_{2}$, and 
${\cal F}$ denotes the fundamental region.
According to the integrand of eq.(\ref{eq:Z_torus}), we find that the density 
distribution function of $\tau$ in the fundamental region is given by up to 
an overall numerical factor
\begin{equation}
\tau_{2}^{-\frac{3+c}{2}} \; e^{-\frac{\pi}{6} \; \tau_{2} \; (1-c)}
|\prod_{n=1}^{\infty} (1- e^{2\pi i \tau n}) |^{2(1-c)}.
\label{eq:moduli_integrand}
\end{equation}
The distribution of $\tau_{2}$ for $c>1$\footnote
{
In the case of the sphere, for example, the anomalous dimensions and the 
string susceptibility turn to be complex for $c>1$.
}
indicates the instability of the vacuum due to the tachyon of the bosonic 
string theory.
It may become a clear evidence for the branched polymers.
%
\section{Numerical results and discussions}
\begin{figure}
\vspace*{0cm}
\centerline{\psfig{file=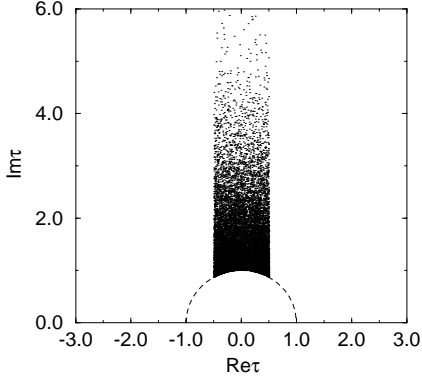,height=5.5cm,width=6.7cm}} 
\vspace{-1.3cm}
\caption
{
Plot of the moduli($\tau$) on the complex-plane with a total number of 
triangles of $16$K.
A dot denotes $\tau$ which is mapped into the fundamental region of each 
configuration 
}
\label{fig:Moduli_Dist_16K}
\vspace{-0.7cm}
\end{figure}

Fig.\ref{fig:Moduli_Dist_16K} shows the distribution of the period $\tau$ 
for a surface with $16$K triangles in the case of the pure gravity 
with about $1.5 \times 10^{4}$ independent configurations.
%
Roughly speaking larger values of $\tau_{2}$ in the fundamental region 
represent tori deformed like a long thin tube, while smaller values of 
$\tau_{2}$ represent almost regular tori.
In order to compare numerical results with the predictions of the 
Liouville theory eq.(\ref{eq:moduli_integrand}), we consider the distribution 
functions integrated over $\tau_{1}$.
%
\begin{figure}
\centerline{\psfig{file=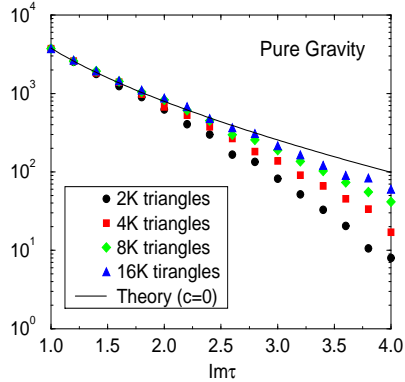,height=5.5cm,width=6cm}} 
\vspace{-1.3cm}
\caption
{
Density-distributions of $\tau_{2}$ in the case of pure-gravity.
}
\label{fig:Hist_Comp_Pure}
\vspace{-0.3cm}
\end{figure}
%
\begin{figure}
\vspace*{-0.3cm}
\centerline{\psfig{file=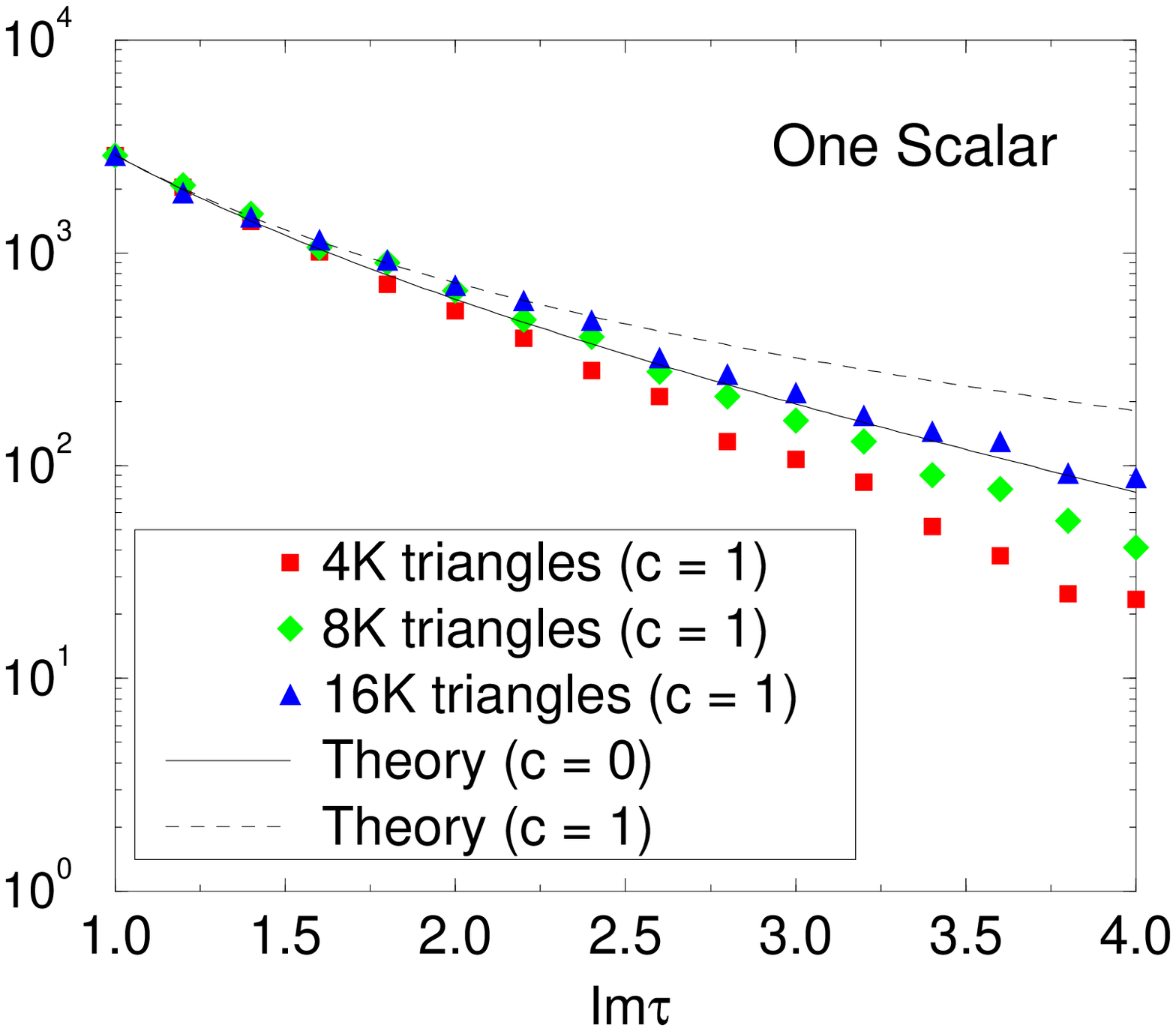,height=5.5cm,width=6cm}} 
\vspace{-1.3cm}
\caption
{
Density-distributions of $\tau_{2}$ in the case of $c=1$.
}
\label{fig:Moduli_Comp_1S}
\end{figure}
%
\begin{figure}
\vspace*{0cm}
\centerline{\psfig{file=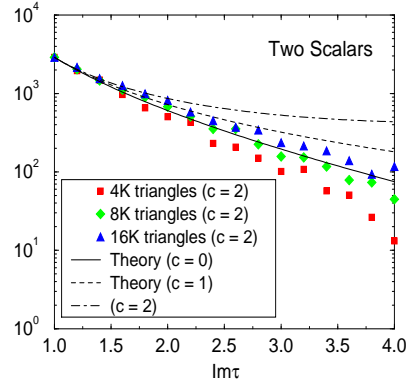,height=5.5cm,width=6cm}} 
\vspace{-1.3cm}
\caption
{
Density-distributions of $\tau_{2}$ in the case of $c=2$.
}
\label{fig:Moduli_Comp_2S}
\vspace{-0.4cm}
\end{figure}
%
Fig.\ref{fig:Hist_Comp_Pure} shows the density distributions of $\tau_{2}$ 
in the case of the pure-gravity with $2$K, $4$K, $8$K and $16$K triangles.
Fig.\ref{fig:Moduli_Comp_1S} shows the distributions of $\tau_{2}$ in 
the case of surfaces coupled with a scalar field($c=1$) with $4$K, $8$K and 
$16$K triangles.
It is clear that the numerical results agree fairly well with the 
predictions of the Liouville theory for sufficiently large number of 
triangles.
Fig.\ref{fig:Moduli_Comp_2S} shows the distributions of $\tau_{2}$ in the 
case of two scalar fields($c=2$) with $4$K, $8$K and $16$K triangles.
In this case, we cannot detect the divergence of the distribution of 
$\tau_{2}$.
It would be hard to obtain a large value of $\tau_{2}$ for relatively small 
number of triangles, because we need many triangles to form a long narrow 
shape to the torus.
We conclude that the DT surfaces have the 
same complex structure as the Liouville theory in the thermodynamic limit 
for $c \leq 1$ cases.
\vspace{-0.3cm}
\begin{center}
-Acknowledgements- 
\end{center}
\vspace{-0.1cm}
\noindent
We are grateful to J.Kamoshita for useful discussions and 
comments.One of the authors(N.T) was supported by a Research Fellowships of 
the Japan Society for the Promotion of Science for Young Scientists.


\end{document}